\documentclass[conference]{IEEEtran}

\usepackage{cite}
\usepackage{amsmath,amssymb,amsfonts}
\usepackage{graphicx}
\usepackage{booktabs}
\usepackage{array}
\usepackage{xcolor}
\usepackage{url}

\graphicspath{{./}{figures/}}

\newcommand{\ABCfam}{\mathrm{ABC\_Family}}
\newcommand{\SMLP}{S40\text{-}MLP}
\newcommand{\SCNN}{S40\text{-}CNN}
\newcommand{\SEMLP}{S80\text{-}MLP}
\newcommand{\SECNN}{S80\text{-}CNN}
\newcommand{\PMLP}{P80\text{-}MLP}
\newcommand{\PCNN}{P80\text{-}CNN}

\begin{document}

\title{The Price of a Cycle: A Phasor -vs- Sampled-Value Comparison for Streaming Fault Classification on the IEEE 9-Bus System}
\author{%
\makebox[\textwidth][c]{%
\begin{tabular}{@{}c@{}}
Emad Abukhousa$^{1}$ \quad
Khaled Muhmmed Nor Muhmmed$^{2}$ \quad
Abdulaziz Qwbaiban$^{2,3}$\\
Saman Zonouz$^{1}$ \quad
A. P. Sakis Meliopoulos$^{1}$\\
$^{1}$\textit{School of Electrical and Computer Engineering, Georgia Institute of Technology, Atlanta, GA, USA}\\
\{emadak, saman.zonouz, sakis.m\}@gatech.edu\\
$^{2}$\textit{Electrical Engineering Department, King Fahd University of Petroleum and Minerals, Dhahran, Saudi Arabia }\\
$^{3}$\textit{Interdisciplinary Research Center for Sustainable Energy Systems (IRC-SES), KFUPM, Dhahran 31261, Saudi Arabia}\\
\{g202518290, qwbaiban\}@kfupm.edu.sa
\end{tabular}%
}
}
\maketitle

\begin{abstract}
Transmission-line fault classification is a time-critical protection function, and recent machine-learning methods classify fault type directly from voltage and current measurements. The input to such a classifier can be supplied as sampled-value windows of raw measurements, which preserve sub-cycle structure, or as phasor-domain sequence components, which are physically interpretable but require a causal one-cycle observation window before a phasor estimate is available. This paper measures the cost of that observation window, the price of a cycle, through a controlled comparison of the two representations for streaming fault-family classification on the standard synchronous WSCC/IEEE 9-bus system. The comparison holds the simulated system, the measured channels, the eleven-class event-family taxonomy, the training procedure, and the streaming decision layer constant, and uses six configurations: 40- and 80-sample sampled-value inputs and one-cycle phasor inputs, each evaluated with a multilayer perceptron (MLP) and a one-dimensional convolutional neural network (CNN). The evaluation reports fault-window accuracy, false-fault and false-normal rates, event-classification success, and the first-correct delay measured from event onset. All six configurations classify every event successfully, yet the sampled-value representation reaches the correct class earlier in every model pair. The 40-sample sampled-value MLP attains the strongest result, with $99.34\%$ fault-window accuracy, $99.79\%$ normal accuracy, a $0.21\%$ false-fault rate, a $0.65\%$ false-normal rate, and a $2.81$~ms mean first-correct delay, and it Pareto-dominates the other five configurations. A matched-window comparison at eighty samples confirms that the timing advantage follows from the representation rather than from the window length. On this clean synchronous benchmark, the one-cycle phasor representation does not return an accuracy or stability benefit sufficient to offset its delay.
\end{abstract}


\section{Introduction}
\label{sec:intro}

Transmission-line fault classification is a foundational protection function, because a protection algorithm must determine not only that a fault exists but which phases are involved and whether the fault is grounded. The decision must be reached quickly enough to support protection-oriented actions such as selective tripping, and to limit equipment stress and unnecessary outages. Machine-learning and deep-learning methods have renewed interest in this function by learning a direct mapping from voltage and current measurements to fault labels, and such methods have been demonstrated on the IEEE 9-bus system with time-series convolutional networks, engineered features with gradient boosting, sequence-component and wavelet features, and dedicated dataset-generation frameworks \cite{Jiriwibhakorn2025IEEE9Bus, Das2025TwelveFaults, Chaudhary2026SequenceWavelet, Kouraichi2026Dataset}. For protection use, however, sample-level accuracy on a static test set is not a sufficient measure of quality, because a classifier that performs well on randomly sampled data may behave differently when replayed as a continuous time stream, and a model with low inference time may still require a measurement or feature-extraction window before any decision can be formed \cite{Oelhaf2025Benchmarking}.

The design choice that governs this measurement window is the input representation, and it is frequently embedded in the data-preparation stage. Sampled-value representations use causal windows of raw voltage and current samples, which preserve the sub-cycle transient structure of a fault and can therefore support a decision shortly after a disturbance begins. Phasor-domain representations instead use estimated sinusoidal quantities, and when those phasors are resolved into symmetrical components, the resulting features carry direct physical meaning for fault analysis \cite{Phadke2009ComputerRelaying}. A causal phasor estimate, however, is not available instantaneously: a one-cycle sliding discrete Fourier transform (DFT) at 60~Hz requires approximately $16.67$~ms of samples before the first phasor is produced, and the transform acts as a windowed spectral estimator whose averaging introduces window-dependent transient response \cite{Oppenheim2010DSP}. The central question of this paper is therefore not which neural architecture classifies faults best, but a representation question: when the same system, channels, event stream, model families, and decision layer are held constant, what does the one-cycle phasor representation cost in timing, and what does it return in accuracy or stability. In this paper, the price of a cycle is the delay introduced when phasor-domain features require one full 60~Hz cycle before the classifier receives a physically meaningful representation, and the study tests whether that delay is justified by improved accuracy or stability compared with direct sampled-value classification.

The relevant literature supports three observations that, taken together, define a specific gap. First, machine-learning fault classification on the IEEE 9-bus system, the twelve-class formulation, and sequence-component features are each individually well-explored, so a contribution cannot rest on any of them in isolation \cite{Jiriwibhakorn2025IEEE9Bus, Das2025TwelveFaults, Chaudhary2026SequenceWavelet, Kouraichi2026Dataset}. Broader transmission-line studies have used extreme-learning-machine and optimized weighted-ELM models, while benchmarking work has emphasized that input-window length and inference time should be reported separately \cite{Goni2023ELMTransmission, Hassan2024OWELM, Oelhaf2025Benchmarking}. Second, phasor and sequence-component representations are physically meaningful for protection, yet causal DFT phasors require a measurement window, many phasor-based classification studies do not explicitly separate the phasor-estimation window from classifier inference time, and phasor-domain approximations can differ from time-domain behavior during transients \cite{Phadke2009ComputerRelaying, Oppenheim2010DSP, Akter2024PMUXAI, Hassani2022TDPhasor}. 
Third, streaming-aware evaluation requires more than static accuracy. Reject-option methods support confidence-gated decisions, and leakage-aware machine-learning practice motivates separating the training stream from an independent replay stream for testing \cite{Chow1970Reject,Geifman2017Selective,Kapoor2023Leakage}. This framing is consistent with the authors' prior work on power-system fault, anomaly, and cyber-physical classification \cite{Abukhousa2025Wisdom,Alsaeed2025Anomaly,Abukhousa2026Latency,Abukhousa2026Transformer,Chen2024PVFaultDiagnosis}. However, those studies did not isolate whether sampled-value windows or one-cycle phasor-domain sequence components are preferable when the phasor-estimation window is counted in the timing budget. This paper addresses that gap. Its novelty is not the benchmark, the taxonomy, or the features in isolation, but the controlled, timing-honest comparison that isolates the input representation as the single experimental variable. 

The contributions are as follows: (i) a controlled comparison of sampled-value and phasor-domain sequence-component representations for streaming fault-family classification on the synchronous IEEE/WSCC 9-bus system, using the same simulated system, measured channels, eleven-class taxonomy, model families, training procedure, and confidence-gated decision layer so that the representation effect can be isolated; (ii) a six-configuration design that separates sub-cycle sampled-value behavior, matched-window sampled-value behavior, and one-cycle phasor behavior across two deep-learning model families, with the model axis used as a robustness control rather than as a contribution; (iii) a timing-honest evaluation that counts the one-cycle phasor-estimation window in the decision timeline and reports first-correct delay from event onset alongside fault-window accuracy, false-fault rate, false-normal rate, and event-classification success; and (iv) a streaming result in which the 40-sample sampled-value multilayer-perceptron configuration Pareto-dominates the other five configurations, with a matched-window comparison establishing that the timing advantage follows from the representation rather than from the window length.

\section{Methodology}
\label{sec:method}

The methodology isolates a single experimental variable, the input representation, while holding the simulated system, the measured channels, the class taxonomy, the training procedure, the model families, and the streaming decision layer constant.

\subsection{Synchronous Test System and Measurement Vector}
\label{subsec:system}
\begin{figure}[!t]
\centering
\includegraphics[width=\columnwidth]{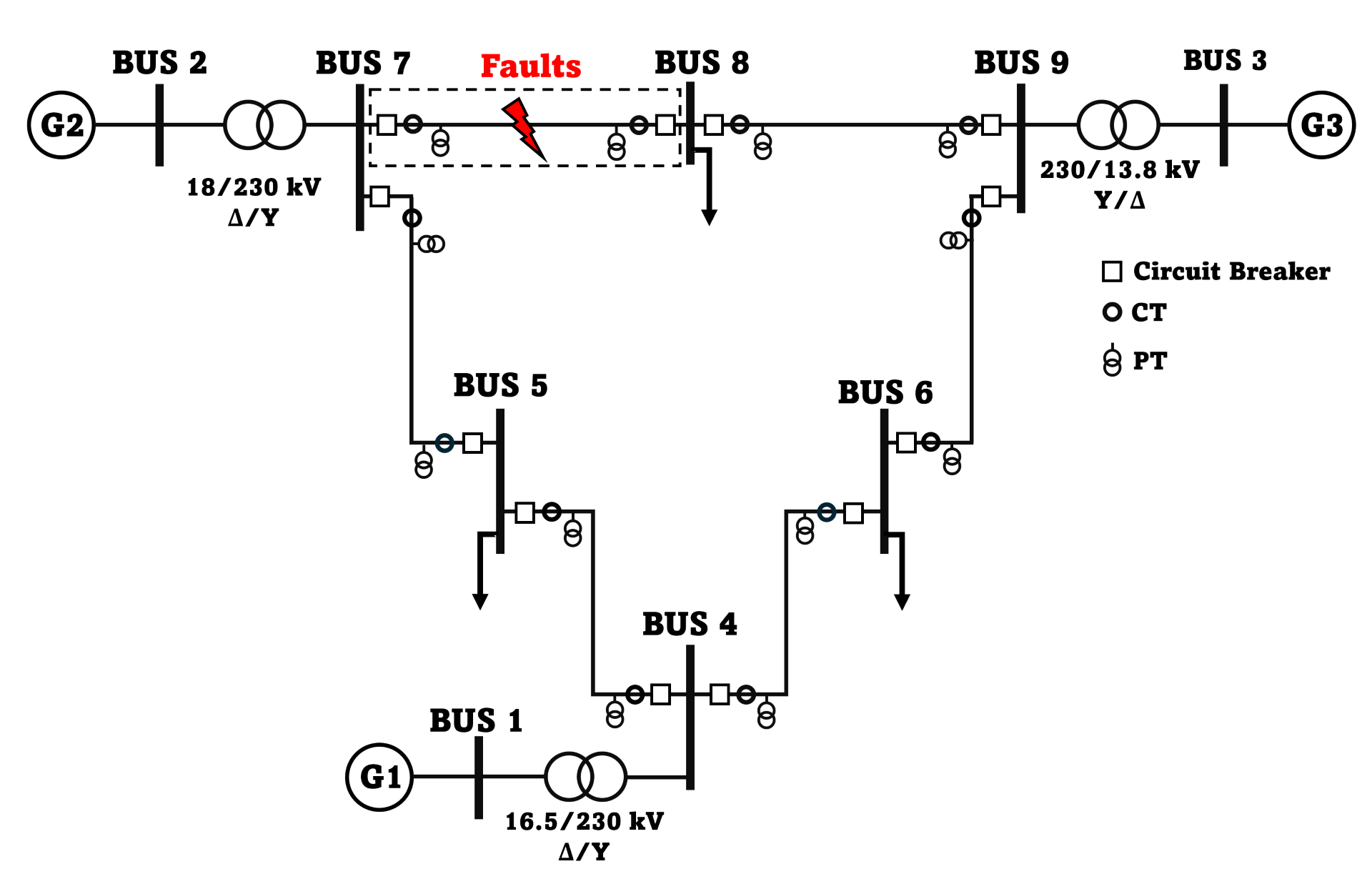}
\caption{Single-line diagram of the synchronous WSCC/IEEE 9-bus test system. Faults are applied at the midpoint of line~7--8. The measured channels are the three-phase voltages at bus~7 and bus~8 and the bidirectional three-phase and neutral line-current channels.}
\label{fig:system}
\end{figure}

The test system is the standard synchronous WSCC/IEEE 9-bus benchmark, simulated offline in RSCAD with a 100~MVA base, three synchronous generators, and three constant-power loads, and with no inverter-based resources. The synchronous configuration is deliberate, because inverter-interfaced resources alter fault-current magnitude, negative-sequence behavior, and protection-zone observability, and their inclusion would confound a study whose sole purpose is to compare two input representations. Faults are applied at the midpoint of line~7--8 with bolted fault resistance, and measurements are taken at both line terminals, as shown in Fig.~\ref{fig:system}.

At sample index $i$, the measurement vector collects the fourteen electrical channels,
\begin{equation}
\label{eq:xvec}
\mathbf{x}_i =
\Big[
V_{7A,i},\, V_{7B,i},\, V_{7C,i},\,
V_{8A,i},\, V_{8B,i},\, V_{8C,i},\,
\end{equation}
\vspace{-1.2em}
\[
I_{78A,i},\, I_{78B,i},\, I_{78C,i},\, I_{78N,i},\,
I_{87A,i},\, I_{87B,i},\, I_{87C,i},\, I_{87N,i}
\Big]^{\top}
\]
namely the three-phase voltages at buses 7 and 8 and the bidirectional three-phase and neutral line currents. The time column is used only for indexing, event alignment, and timing; it is never a model feature. The class label is used only for training and validation and is absent from the blind replay stream.

\subsection{Data Streams and Scoring Windows}
\label{subsec:data}
The study uses two physically separate streams generated from the same fault configuration. The labeled training stream contains normal operation together with eleven one-second bolted midpoint faults, namely AG, BG, CG, AB, BC, CA, ABG, BCG, CAG, ABC, and ABCG. For the active eleven-class event-family task, ABC and ABCG are mapped to a single $\ABCfam$ class, as defined in \eqref{eq:remap}. The training stream supports model fitting, scaler estimation, and validation.

The testing replay uses the same fault families as training but with a different event order, different event start times, and different event durations. The test event times and durations are intentionally different from the training stream so that the classifier is not evaluated only at the same event positions and one-second durations used for training. This design tests temporal robustness, onset invariance, and duration robustness under the same fixed electrical condition, and it does not test generalization to unseen locations, resistances, topologies, or loading conditions. Normal intervals are placed between consecutive events so that post-fault clearing and return-to-normal behavior are visible in the stream and can be evaluated for false-fault behavior. The event windows used for scoring and for figure shading are defined in stream time and are listed in Table~\ref{tab:test_events}.

The sampled-value stream follows an IEC~61850-style sampled-value rate of $80$ samples per $60$~Hz cycle. The nominal sampling rate is therefore $4.8$~kHz and the sampling interval is $208.33~\mu$s. One cycle at $60$~Hz is $1/60~\mathrm{s} = 16.67$~ms and corresponds to $80$ samples, while a half cycle is $1/120~\mathrm{s} = 8.33$~ms and corresponds to $40$ samples. The one-cycle phasor estimator therefore uses $N = 80$ samples, and the $40$-sample sampled-value window is the half-cycle representation rather than an arbitrary choice.

\begin{table}[!t]
\centering
\caption{Testing replay event windows used for streaming evaluation and figure shading.}
\label{tab:test_events}
\begin{tabular}{c c l c c}
\toprule
Event & Class & Fault & Start (s) & End (s) \\
\midrule
1  & 10 & ABC\_Family (ABCG) & 1.026528  & 1.326672 \\
2  & 2  & BG                 & 3.326592  & 3.776704 \\
3  & 9  & CAG                & 5.776624  & 6.026640 \\
4  & 4  & AB                 & 8.026560  & 8.526592 \\
5  & 3  & CG                 & 10.526720 & 10.876576 \\
6  & 6  & CA                 & 12.876704 & 13.476576 \\
7  & 10 & ABC\_Family (ABC)  & 15.476704 & 15.776640 \\
8  & 5  & BC                 & 17.776560 & 18.226672 \\
9  & 8  & BCG                & 20.226592 & 20.626576 \\
10 & 1  & AG                 & 22.626704 & 23.176656 \\
11 & 7  & ABG                & 25.176576 & 25.426592 \\
\bottomrule
\end{tabular}
\end{table}

\subsection{Sampled-Value and Phasor Representations}
\label{subsec:repr}
The two representations differ only in how the fourteen channels are presented to an identical classifier, and both are causal. For window length $L$, the causal sampled-value input stacks the most recent $L$ measurement vectors,
\begin{equation}
\label{eq:swin}
\mathbf{X}_{i,L} = \big[\,\mathbf{x}_{i-L+1}\;\; \cdots\;\; \mathbf{x}_{i}\,\big]^{\!\top} \in \mathbb{R}^{L \times 14}
\end{equation}
The 40-sample window spans $T_{40} = 1/120~\mathrm{s} \approx 8.33$~ms, which is sub-cycle, and the 80-sample window spans $T_{80} \approx 16.67$~ms, which matches the one-cycle phasor window and serves as a matched-window control. Each channel is standardized using training-split statistics $\mu_d,\sigma_d$,
\begin{equation}
\label{eq:zscore}
z_{i,d} = \frac{x_{i,d} - \mu_d}{\sigma_d + \epsilon}
\end{equation}
and the same statistics are reused for validation and replay, which prevents leakage into the standardization. Each training window receives the label of its final sample, preserving causality.

The phasor representation estimates the fundamental component of each channel from the trailing one-cycle window. For a real-valued channel $x[n]$, the causal one-cycle DFT phasor is
\begin{equation}
\label{eq:dft}
X_i = \frac{2}{N}\sum_{m=0}^{N-1} x[i-N+1+m]\,e^{-j2\pi m/N}
\end{equation}
with $N=80$ and a peak-amplitude convention; a centered window is inadmissible because it would require future samples. The three-phase phasors are resolved into symmetrical components with the Fortescue operator $a = e^{j2\pi/3}$,
\begin{equation}
\label{eq:seq}
\begin{bmatrix} X_0 \\ X_1 \\ X_2 \end{bmatrix}
= \frac{1}{3}
\begin{bmatrix} 1 & 1 & 1 \\ 1 & a & a^{2} \\ 1 & a^{2} & a \end{bmatrix}
\begin{bmatrix} X_A \\ X_B \\ X_C \end{bmatrix}
\end{equation}
which separates fault types by physics, since single-line-to-ground faults excite zero- and negative-sequence content, line-to-line faults emphasize the negative sequence, double-line-to-ground faults excite both, and balanced three-phase faults are dominated by the positive sequence. The phasor feature vector concatenates the real and imaginary parts of the sequence components for the four three-phase groups $V7,V8,I78,I87$ together with the two neutral-current phasors, giving dimension $4(3)(2)+2(2)=28$.

The classifier targets an eleven-class event-family taxonomy. The original twelve labels comprise normal operation, three single-line-to-ground, three line-to-line, three double-line-to-ground, and the two three-phase faults ABC and ABCG. The two three-phase labels are merged into a single \mbox{$\ABCfam$} class, because a bolted three-phase fault carries near-zero residual current and the ground/no-ground distinction is physically marginal for fast event-family classification,
\begin{equation}
\label{eq:remap}
g(y) = \begin{cases} 10, & y \in \{10, 11\},\\ y, & y \in \{0,\dots,9\} \end{cases}
\end{equation}

\subsection{Model Families and Training}
\label{subsec:models}
Two model families, a multilayer perceptron (MLP) and a one-dimensional convolutional network (CNN), serve as architecture controls so that representation trends can be observed across a model without temporal structure and a model with it. Both produce logits $\mathbf{z}_i \in \mathbb{R}^{K}$ with $K=11$, followed by a softmax. The sampled-value MLP flattens \eqref{eq:swin} and applies dense layers of width $256$, $128$, and $64$ with batch normalization, rectified-linear activation, and dropout $0.2$; the phasor MLP applies dense layers of width $128$ and $64$ to the 28-dimensional vector. The sampled-value CNN treats the fourteen channels as input channels and the sample index as the temporal axis, applying convolutions of $32$, $64$, and $64$ filters with kernel lengths $3$, $5$, and $5$, followed by global average pooling and a dense classifier. The phasor CNN arranges the 28 features as a two-channel real/imaginary representation over the ordered feature axis and convolves along that axis, which exposes the phasor features to a convolutional inductive bias rather than a temporal convolution. Because normal operation dominates the stream, training uses a class-weighted cross-entropy with weight $w_c = N/(K N_c)$, optimized by AdamW at learning rate $10^{-3}$ and weight decay $10^{-4}$, batch size $256$, at most $50$ epochs, and early stopping on the validation macro-$F_1$.

\subsection{Streaming Decision Layer and Metrics}
\label{subsec:stream}
The streaming layer converts per-sample probabilities into an emitted decision through causal smoothing, confidence gating, and optional persistence. A causal trailing moving average over $W$ samples,
\begin{equation}
\label{eq:smooth}
\bar{p}_{i,k} = \frac{1}{|\mathcal{W}_i|}\!\!\sum_{r=\max(0,\,i-W+1)}^{i}\!\! p_{r,k}
\end{equation}
is used rather than a centered window, which would leak future samples. With confidence $c_i = \max_k \bar{p}_{i,k}$, smoothed prediction $\tilde{y}_i = \arg\max_k \bar{p}_{i,k}$, and threshold $\tau=0.60$, the emitted class is
\begin{equation}
\label{eq:gate}
e_i = \begin{cases} \tilde{y}_i, & c_i \ge \tau,\\ -1, & c_i < \tau, \end{cases}
\end{equation}
where $-1$ denotes abstention \cite{Chow1970Reject, Geifman2017Selective}. The gate and an optional persistence rule are retained even though the clean benchmark produces confident decisions throughout, so that the same pipeline transfers to noisier settings. For event $e$ with index set $\mathcal{I}_e=\{i:s_e\le t_i<q_e\}$ and true class $y_e$, the event-window accuracy is $A_e = |\mathcal{I}_e|^{-1}\sum_{i\in\mathcal{I}_e}\mathbb{1}(e_i=y_e)$, and the aggregate fault-window accuracy pools all fault-window samples. The first-correct delay is
\begin{equation}
\label{eq:delay}
D_e^{\mathrm{correct}} = 1000\big(t_e^{\mathrm{fc}} - s_e\big)~\text{ms}, \quad
t_e^{\mathrm{fc}} = \min\{t_i : e_i = y_e,\, i\in\mathcal{I}_e\}
\end{equation}
The first-correct delay can be shorter than the window length of \eqref{eq:swin}, because the classifier emits a decision at every sample and a trailing window accumulates an increasing fraction of post-onset samples, so a correct decision can be reached before the window is filled entirely with post-onset data.

\section{Results}
\label{sec:results}

The evaluation measures the streaming behavior of the six configurations on the unlabeled replay stream, under the common decision layer and event-window scoring of Section~\ref{subsec:stream}. The central question is timing-sensitive: once a fault begins, which representation and model family reaches the correct class earliest, and what fault-window accuracy does it sustain.

\subsection{Model-Level Performance}
\label{subsec:res-overall}
All six configurations classify every event successfully, yet they separate sharply on accuracy and on the time required to reach a correct decision. Each configuration attains an event classification success rate of $11/11$, so representation and architecture do not determine whether a fault is eventually identified; the configurations separate instead on fault-window accuracy, false-normal rate, and first-correct delay (Table~\ref{tab:model}). In the configuration names, S40 and S80 denote 40- and 80-sample sampled-value windows and P80 denotes the 80-sample one-cycle phasor representation, while MLP and CNN denote the two model families.

\begin{table*}[!t]
\centering
\caption{Streaming model-level results on the eleven test events.}
\label{tab:model}
\begin{tabular}{l c c c c c c c c c}
\toprule
Configuration & Repr. & Model & Window & Fault acc. & Normal acc. & FFR & FNR & Mean delay & Median delay \\
 & & & samples & (\%) & (\%) & (\%) & (\%) & (ms) & (ms) \\
\midrule
$\SMLP$  & sampled & MLP & 40 & \textbf{99.338} & \textbf{99.792} & \textbf{0.208} & \textbf{0.652} & \textbf{2.807} & \textbf{2.656} \\
$\SCNN$  & sampled & CNN & 40 & 97.844 & 99.728 & 0.272 & 1.536 & 3.583 & 3.280 \\
$\SEMLP$ & sampled & MLP & 80 & 99.159 & 99.767 & 0.233 & 0.799 & 3.526 & 3.280 \\
$\SECNN$ & sampled & CNN & 80 & 99.022 & 99.705 & 0.295 & 0.964 & 4.074 & 3.904 \\
$\PMLP$  & phasor  & MLP & 80 & 98.639 & 99.534 & 0.466 & 1.347 & 5.606 & 5.360 \\
$\PCNN$  & phasor  & CNN & 80 & 97.963 & 99.587 & 0.413 & 1.811 & 7.270 & 6.400 \\
\bottomrule
\end{tabular}
\end{table*}

The 40-sample sampled-value MLP leads on every reported axis, reaching $99.338\%$ fault-window accuracy and $99.792\%$ normal accuracy, with a $0.208\%$ false-fault rate, a $0.652\%$ false-normal rate, and a $2.807$~ms mean first-correct delay. A single configuration therefore leads on both objectives at once, which is uncommon, because the most accurate model in a classification study is frequently the slowest. The phasor configurations classify all events but transition to the correct class later than their sampled-value counterparts: the phasor MLP attains $98.639\%$ accuracy with a $5.606$~ms delay, and the phasor CNN attains $97.963\%$ accuracy with a $7.270$~ms delay and the largest false-normal rate, $1.811\%$. The phasor representation thus does not fail; it transitions more slowly under the present clean conditions.

The window-length comparison within the MLP family shows that the sub-cycle window suffices, because extending the sampled-value MLP from 40 to 80 samples lowers fault-window accuracy from $99.338\%$ to $99.159\%$ and raises the mean delay from $2.807$~ms to $3.526$~ms, so the longer window adds delay without an accuracy gain. The architecture comparison is consistent in timing but mixed in accuracy. Within each representation the sampled-value configuration is faster, so the timing advantage holds across both families; the accuracy ordering, however, depends on architecture and window length, because the 40-sample sampled-value CNN is faster than the phasor CNN yet slightly less accurate ($97.844\%$ against $97.963\%$), whereas the 80-sample sampled-value CNN is both faster and more accurate. The defensible statement is therefore that sampled-value inputs yield consistently earlier decisions, while their accuracy advantage is clearest for the MLP family and is window-dependent for the CNN family.

\subsection{Matched-Window Comparison}
\label{subsec:res-matched}
The matched-window comparison separates the representation effect from the window-length effect by fixing the observation window at 80 samples for both arms. Comparing $\SEMLP$ against $\PMLP$, and $\SECNN$ against $\PCNN$, varies only the representation, and in both matched pairs the sampled-value configuration records the lower delay and the higher fault-window accuracy (Table~\ref{tab:model}). The timing advantage of the sampled-value representation is therefore not an artifact of the shorter 40-sample window; the representation itself governs how quickly the streaming decision settles after onset.

\subsection{Full-Stream Behavior}
\label{subsec:res-fullstream}

\begin{figure}[!t]
\centering
\includegraphics[width=\columnwidth]{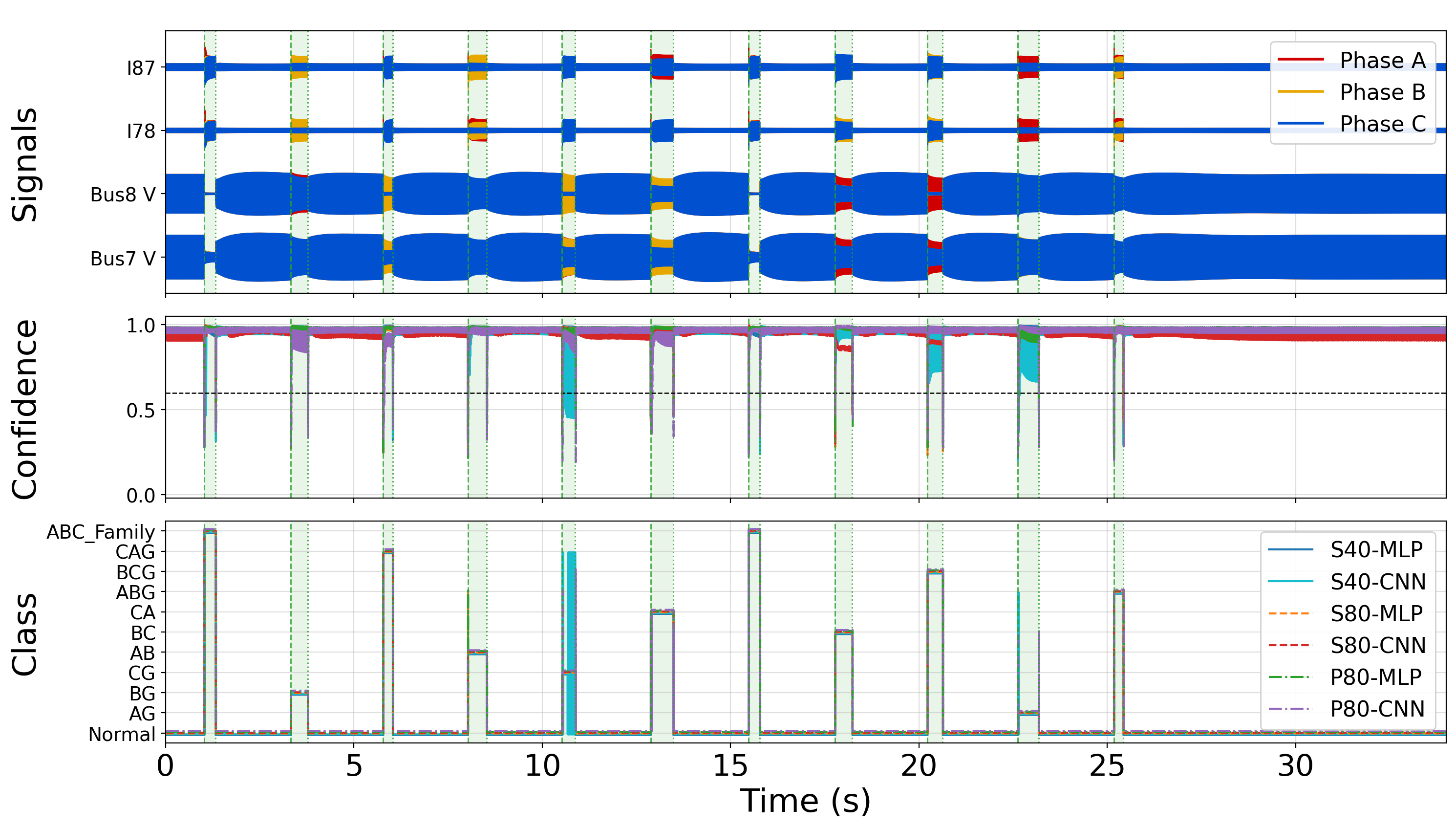}
\caption{Full-stream overlay of the testing replay. The shaded regions mark the eleven event windows used for scoring. The upper panel shows selected voltage and current channels, the middle panel shows model confidence with the gate at $0.60$, and the lower panel shows the emitted class of all six configurations.}
\label{fig:fullstream}
\end{figure}

The full-stream overlay verifies that the reported model-level metrics are produced by sustained streaming behavior across the entire replay, not by isolated event samples (Fig.~\ref{fig:fullstream}). The shaded regions in the figure are the testing replay event windows of Table~\ref{tab:test_events}. Within these intervals, the emitted class moves from Normal to the corresponding fault family after event onset, while between intervals the emitted class returns to Normal after clearing. The alignment between the shaded windows and the visible voltage/current disturbances confirms that the evaluation measures the response to the physical event intervals used for scoring.

The confidence traces provide the link between the raw electrical response and the emitted class decisions. Confidence remains high during most normal and fault intervals, but it decreases near event onsets and clearings, where the causal window contains a mixture of pre-event and event samples or event and post-event samples. These boundary regions are where the confidence gate and possible abstention behavior are most relevant. The low false-fault rate of the best configuration, $0.208\%$, further confirms that the models do not generate frequent false indications during the normal intervals between events.
\subsection{Accuracy-Delay Tradeoff}
\label{subsec:res-pareto}

\begin{figure}[!t]
\centering
\includegraphics[width=\columnwidth]{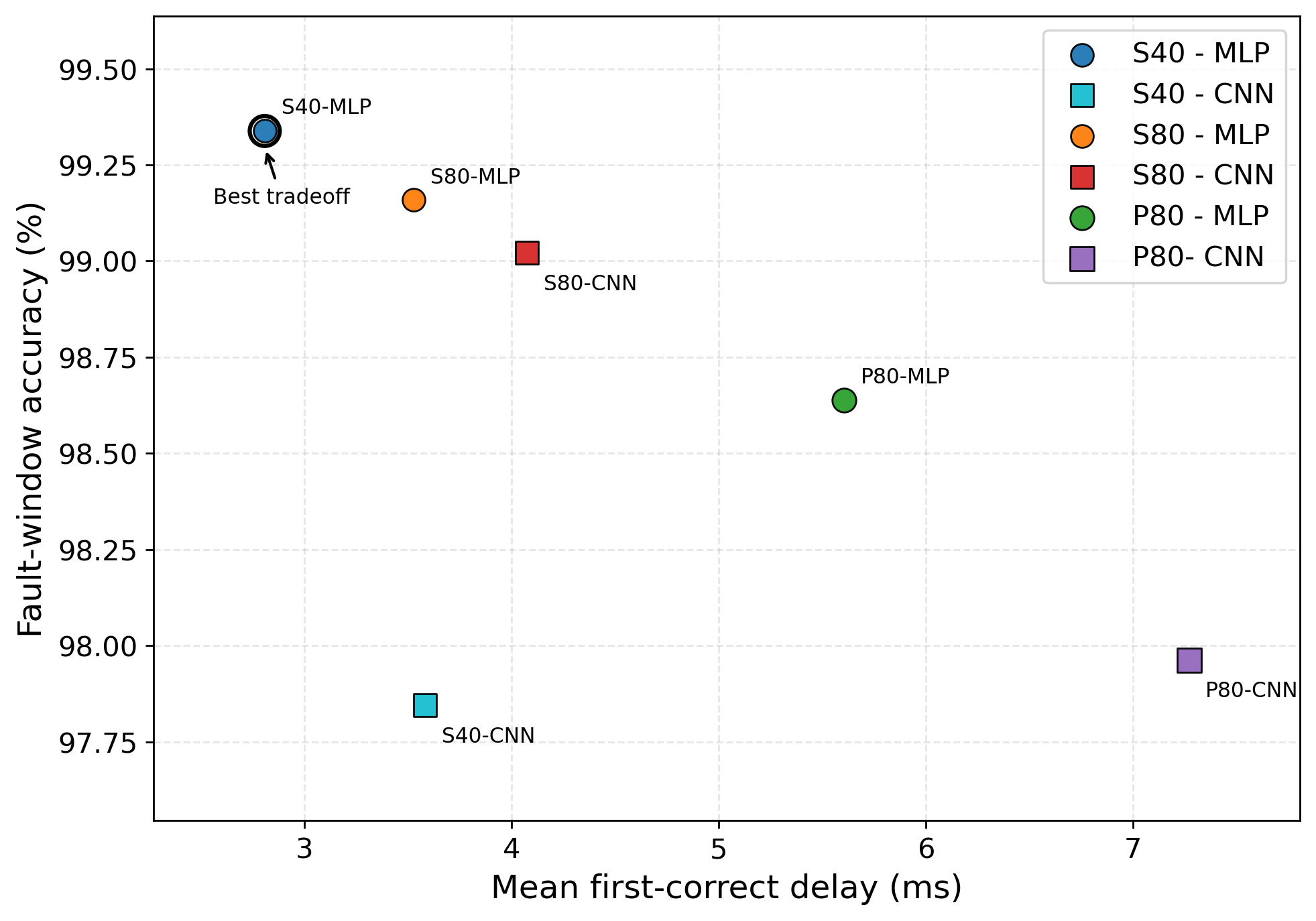}
\caption{Accuracy-delay tradeoff across the six configurations. The horizontal axis is the mean first-correct delay and the vertical axis is the fault-window accuracy, so the upper-left region is preferred. The 40-sample sampled-value MLP combines the highest accuracy with the shortest delay and Pareto-dominates the other five configurations.}
\label{fig:pareto}
\end{figure}

The same streaming results can also be viewed from an accuracy-delay perspective, which reduces the comparison to two protection-oriented objectives: high fault-window accuracy and low first-correct delay (Fig.~\ref{fig:pareto}). The full-stream overlay shows how the configurations respond over time, whereas the accuracy-delay plot summarizes the resulting performance coordinates for each configuration. The sampled-value configurations occupy lower-delay regions than the phasor configurations, and $\SMLP$ occupies the region of highest accuracy and lowest delay. Under the Pareto criterion, in which configuration $a$ dominates configuration $b$ when $A_a \ge A_b$ and $\bar{D}_a \le \bar{D}_b$ with at least one strict inequality, $\SMLP$ dominates the other five configurations. The tradeoff frontier therefore collapses to a single configuration in the tested set.

The two axes do not contribute equally to the interpretation. The fault-window accuracy differences are real but modest, since $\SMLP$ exceeds $\PMLP$ by $0.699$ percentage points. The first-correct delay differences are larger relative to the sub-cycle time scale: the phasor MLP requires $2.799$~ms more than $\SMLP$, and the phasor CNN requires $3.687$~ms more than $\SCNN$. The defensible reading is therefore that the sampled-value representation provides consistently earlier correct decisions while maintaining comparable event-family accuracy, with timing, rather than accuracy, as the dominant difference.

\subsection{Event Case Studies}
\label{subsec:res-event}
\begin{figure}[!t]
\centering
\includegraphics[width=\columnwidth]{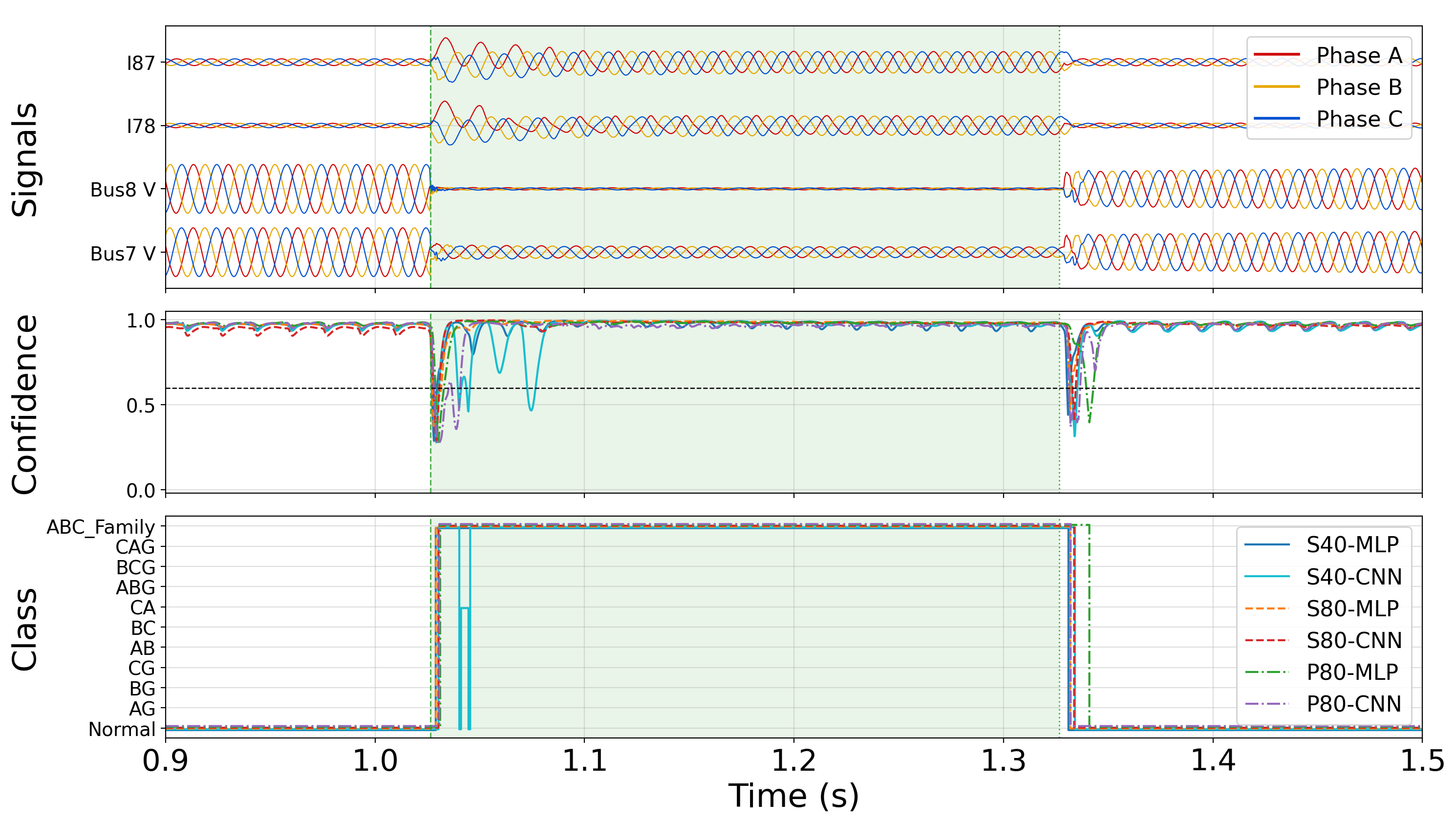}
\caption{Event~1, an $\ABCfam$ three-phase-family fault over $0.90$--$1.50$~s. All six configurations enter the three-phase-family class shortly after onset.}
\label{fig:event1}
\end{figure}

\begin{figure}[!t]
\centering
\includegraphics[width=\columnwidth]{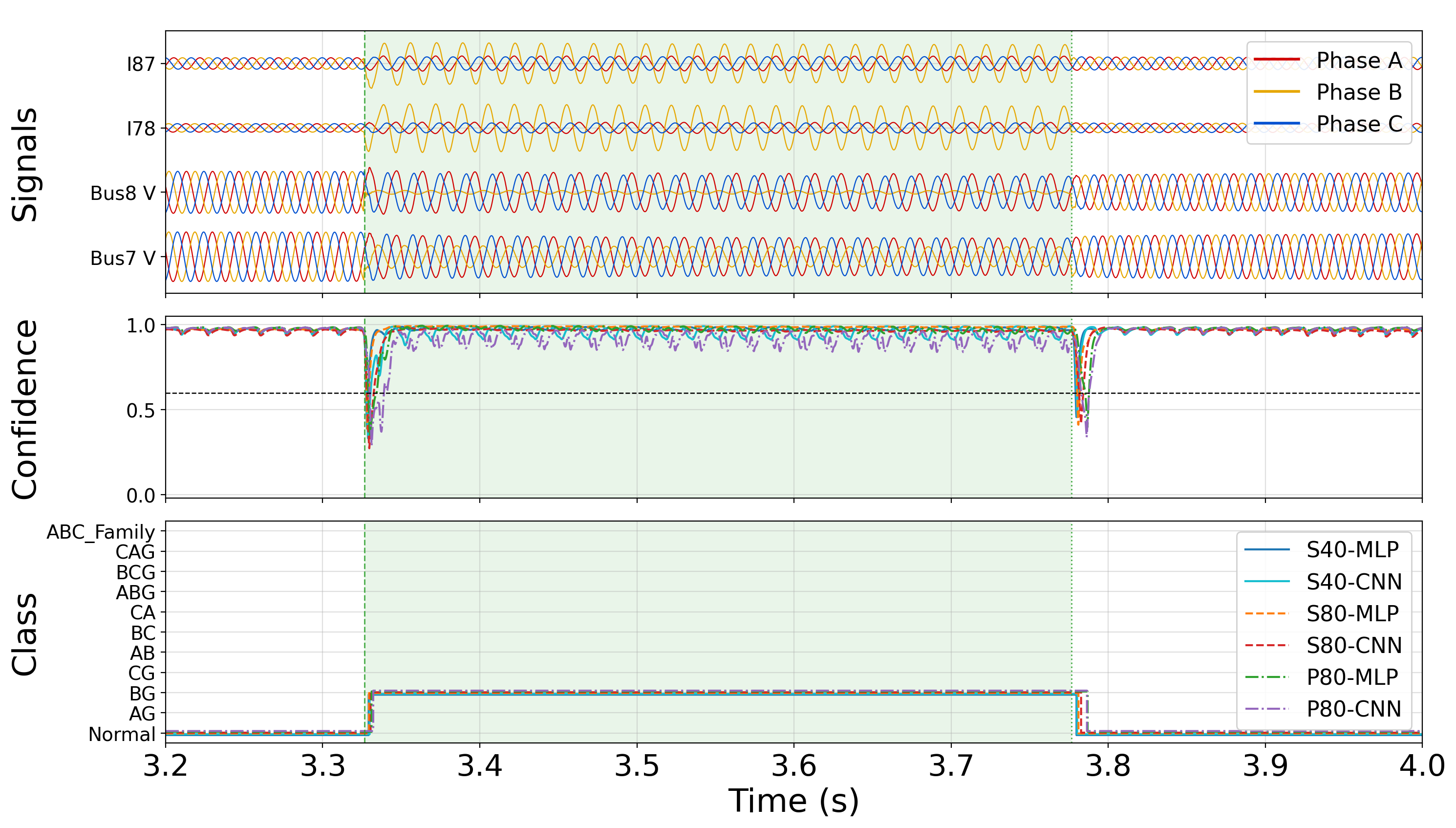}
\caption{Event~2, the BG single-line-to-ground fault over $3.200$--$4.000$~s. All six configurations transition from normal to the BG family inside the event window, and the phasor traces show slower, more oscillatory confidence than the sampled-value traces.}
\label{fig:event2}
\end{figure}

\begin{figure}[!t]
\centering
\includegraphics[width=\columnwidth]{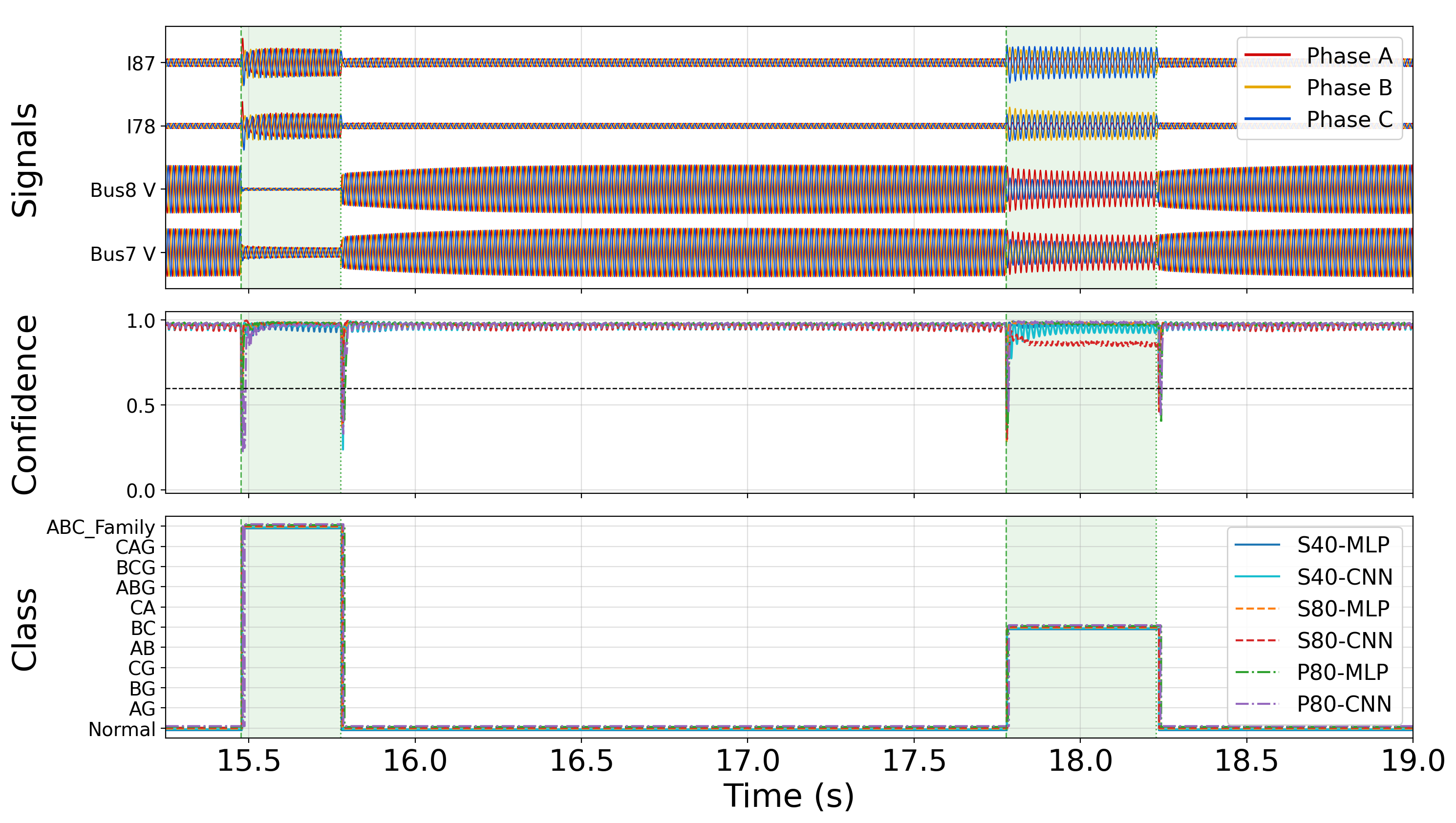}
\caption{Multi-event window over $15.25$--$19.00$~s, showing two successive fault events and the classification response of multiple models.}
\label{fig:eventwide}
\end{figure}

Selected events isolate the detection and classification behavior in detail. The first $\ABCfam$ event shows all six configurations entering the three-phase-family class shortly after onset, which confirms that the $\ABCfam$ grouping classifies a three-phase event as one family rather than splitting it between ABC and ABCG (Fig.~\ref{fig:event1}). The BG event shows the configurations assigning the specific single-line-to-ground family rather than a generic fault indication, and the phasor traces exhibit slower and more oscillatory confidence than the fastest sampled-value traces (Fig.~\ref{fig:event2}). A wider multi-event window shows that the relative ordering of the configurations is consistent from one fault family to the next (Fig.~\ref{fig:eventwide}).

\begin{figure}[!t]
\centering
\includegraphics[width=\columnwidth]{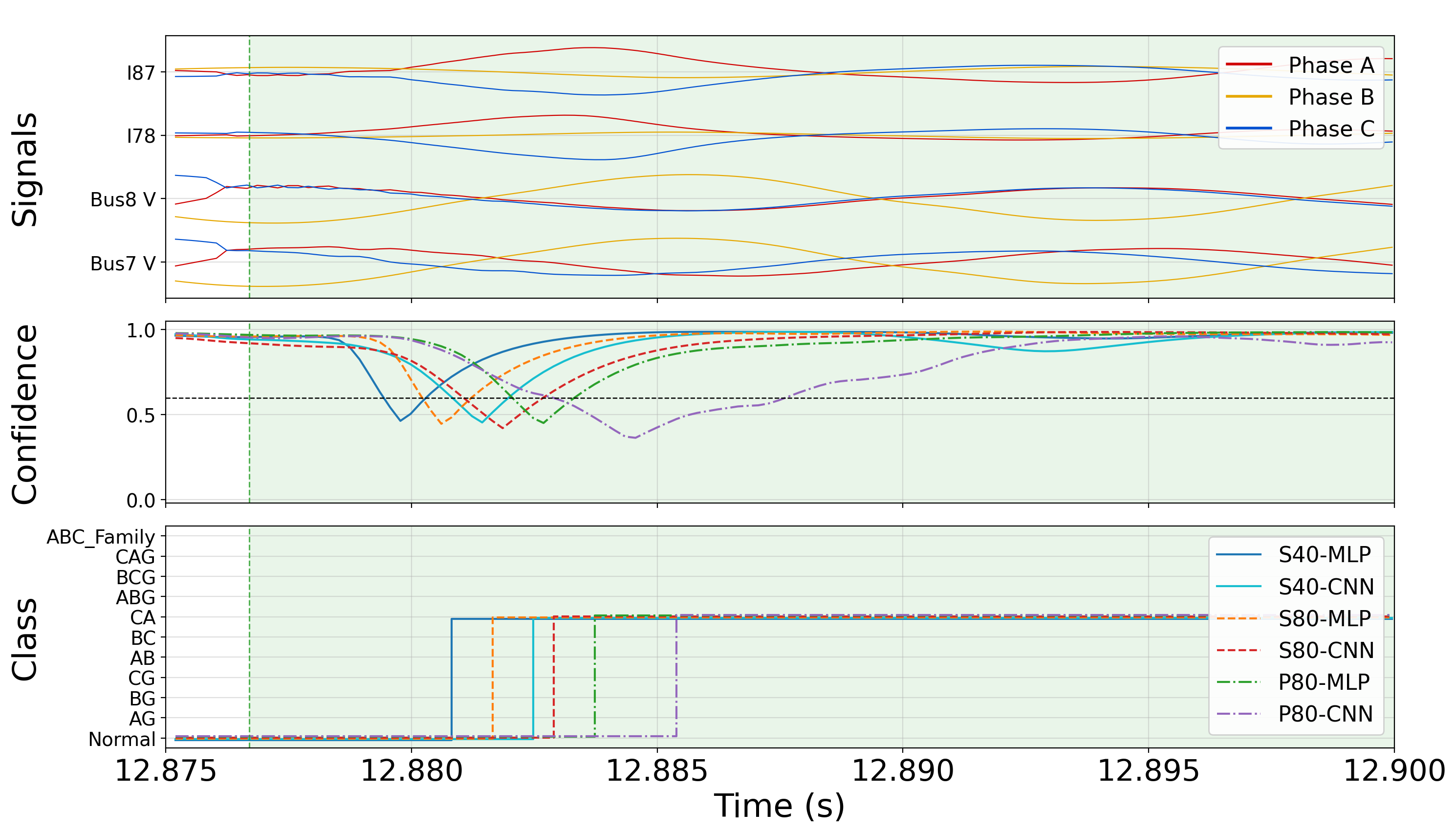}
\caption{Zoomed view of the CA line-to-line fault onset over $12.875$--$12.900$~s. The CA class represents a fault between phase C and phase A. The lower panel shows the order in which the six configurations transition from Normal to the CA fault class.}
\label{fig:caonset}
\end{figure}

The onset comparison provides the physical counterpart to the tradeoff figure. A zoomed view of the CA line-to-line fault onset, involving phases C and A, shows the emitted class of each configuration transitioning from Normal to the CA fault class in sequence, with $\SMLP$ first, the remaining sampled-value configurations following, and the phasor configurations last (Fig.~\ref{fig:caonset}). The operating points in the accuracy-delay figure therefore arise from the time-domain transition behavior observed at event onset. This event-level view also exposes the architecture-dependent stability noted above: the 40-sample sampled-value CNN detects and classifies the event, but its confidence and class trace exhibit more fluctuation than the corresponding MLP response. This behavior helps explain why fault-window accuracy and false-normal rate, rather than event success alone, are necessary to characterize streaming performance.

\section{Discussion}
\label{sec:discussion}

The price of a cycle, in this benchmark, is a timing cost rather than a classification failure. The phasor arm supplies physically structured sequence components and classifies every event, but its one-cycle estimation window delays the first correct decision, and the clean replay does not reveal a compensating accuracy or stability benefit. The result is therefore narrow and should be stated precisely: under fixed-location bolted faults on a synchronous benchmark, sampled-value inputs provide consistently earlier correct decisions with comparable event-family accuracy, and the 40-sample sampled-value MLP Pareto-dominates the other five configurations. The accuracy advantage is clearest for the MLP family, while the CNN accuracy ordering is mixed, so the universal claim is the timing claim, not an accuracy claim.

Two interpretive points follow. First, the first-correct delays are shorter than the nominal window lengths because the classifier emits a trailing-window decision at every sample, so a correct decision can be reached before the window fills entirely with post-onset data; the delay is an empirical streaming-transition quantity rather than the representation-window duration. Second, the gradual phasor transition that delays the decision here may become advantageous elsewhere, because the one-cycle DFT averages over a full cycle and can suppress noise, harmonics, and short disturbances, so the phasor representation may recover value under measurement noise, missing channels, or broader operating conditions that are absent from the present clean benchmark.

The scope of the conclusions is bounded by the experimental design. The study fixes the fault location at the midpoint of line~7--8, uses bolted fault resistance, and holds the operating condition fixed; it does not include a noise sweep, faults outside the protection zone, or variation in load level and generator dispatch, and it does not report hardware-in-the-loop timing. The taxonomy also groups ABC and ABCG into a single $\ABCfam$ class rather than resolving three-phase ground involvement. These boundaries identify the natural extensions: a controlled signal-to-noise sweep that would test whether the phasor representation recovers value, a sampled-value window-length sweep from one to eighty samples to characterize the full sub-cycle-to-cycle tradeoff, a fault-location sweep along and beyond the protected line, a fault-resistance sweep, out-of-zone faults, load and generator-dispatch variation, channel-masking ablations, and hardware-in-the-loop timing, followed by transfer of the same comparison framework to inverter-rich and cyber-physical testbeds.

\section{Conclusion}
\label{sec:conclusion}

This paper presented a controlled, timing-honest comparison of sampled-value and phasor-domain sequence-component representations for streaming fault-family classification on the synchronous IEEE/WSCC 9-bus system, holding the system, channels, eleven-class taxonomy, model families, training procedure, and streaming decision layer constant so that the representation was the only variable. Across six configurations, all classifiers identified every event, but the sampled-value representation reached the correct class earlier in every model pair. The 40-sample sampled-value MLP achieved $99.34\%$ fault-window accuracy, $99.79\%$ normal accuracy, a $0.21\%$ false-fault rate, a $0.65\%$ false-normal rate, and a $2.81$~ms mean first-correct delay, and it Pareto-dominated the other five configurations; a matched-window comparison established that the timing advantage follows from the representation rather than from the window length. On this clean synchronous benchmark, the one-cycle phasor representation did not return an accuracy or stability benefit sufficient to offset its delay, although it may do so under noise or broader operating conditions, which the outlined extensions will test.

\bibliographystyle{IEEEtran}
\bibliography{references}

\end{document}